\documentclass[aps,pra,twocolumn,letterpaper,superscriptaddress,10pt]{revtex4-1} 
\usepackage{amssymb,amsthm,amsmath,amsfonts}
\usepackage{graphicx,mathptmx}
\usepackage{upgreek}
\usepackage{textcomp}
\usepackage{amsmath}
\usepackage{array, multirow, bigdelim, makecell, booktabs}
\usepackage{color}
\usepackage{soul}
\usepackage{physics}
\usepackage{mathtools}

\usepackage[shortlabels]{enumitem}
\usepackage[pdftex,dvipsnames,usenames]{xcolor}
\usepackage[colorlinks=true,urlcolor=blue,citecolor=blue,linkcolor=blue]{hyperref}
\usepackage{gensymb}

\begin{document}

\title{Simultaneous transmission of hyper-entanglement in 3 degrees of freedom through a multicore fiber}

\author{Lukas Achatz}
	\email{lukas.achatz@oeaw.ac.at}
	\affiliation{Institute for Quantum Optics and Quantum Information - IQOQI Vienna, Austrian Academy of Sciences, Boltzmanngasse 3, 1090 Vienna, Austria}
	\affiliation{Vienna Center for Quantum Science and Technology (VCQ), Vienna, Austria}
\author{Lukas Bulla}
	\affiliation{Institute for Quantum Optics and Quantum Information - IQOQI Vienna, Austrian Academy of Sciences, Boltzmanngasse 3, 1090 Vienna, Austria}
    \affiliation{Vienna Center for Quantum Science and Technology (VCQ), Vienna, Austria}
\author{Evelyn A. Ortega}
	\affiliation{Institute for Quantum Optics and Quantum Information - IQOQI Vienna, Austrian Academy of Sciences, Boltzmanngasse 3, 1090 Vienna, Austria}
	\affiliation{Vienna Center for Quantum Science and Technology (VCQ), Vienna, Austria}
\author{Michael Bartokos}
    \affiliation{Institute for Quantum Optics and Quantum Information - IQOQI Vienna, Austrian Academy of Sciences, Boltzmanngasse 3, 1090 Vienna, Austria}
\author{Sebastian Ecker}
	\affiliation{Institute for Quantum Optics and Quantum Information - IQOQI Vienna, Austrian Academy of Sciences, Boltzmanngasse 3, 1090 Vienna, Austria}
	\affiliation{Vienna Center for Quantum Science and Technology (VCQ), Vienna, Austria}
\author{Martin Bohmann}
    \affiliation{Institute for Quantum Optics and Quantum Information - IQOQI Vienna, Austrian Academy of Sciences, Boltzmanngasse 3, 1090 Vienna, Austria}
	\affiliation{Vienna Center for Quantum Science and Technology (VCQ), Vienna, Austria}
\author{Rupert Ursin}
	\affiliation{Institute for Quantum Optics and Quantum Information - IQOQI Vienna, Austrian Academy of Sciences, Boltzmanngasse 3, 1090 Vienna, Austria}
	\affiliation{Vienna Center for Quantum Science and Technology (VCQ), Vienna, Austria}
\author{Marcus Huber}
    \email{marcus.huber@oeaw.ac.at}
	\affiliation{Institute for Quantum Optics and Quantum Information - IQOQI Vienna, Austrian Academy of Sciences, Boltzmanngasse 3, 1090 Vienna, Austria}
	\affiliation{Vienna Center for Quantum Science and Technology (VCQ), Vienna, Austria}

\begin{abstract}
Entanglement distribution is at the heart of most quantum communication protocols. Inevitable loss of photons along quantum channels is a major obstacle for distributing entangled photons over long distances, as the no-cloning theorem forbids the information to simply be amplified along the way as is done in classical communication. It is therefore desirable for every successfully transmitted photon pair to carry as much entanglement as possible. Spontaneous parametric down-conversion (SPDC) creates photons entangled in multiple high-dimensional degrees of freedom simultaneously, often referred to as hyper-entanglement. In this work, we use a multicore fibre (MCF) to show that energy-time and polarization degrees of freedom can simultaneously be transmitted in multiple fibre cores, even maintaining path entanglement across the cores. We verify a fidelity to the ideal Bell state of at least 95$\%$ in all degrees of freedom. Furthermore, because the entangled photons are created with a center wavelength of 1560 nm, our approach can readily be integrated into modern telecommunication infrastructure, thus paving the way for high-rate quantum key distribution and many other entanglement-based quantum communication protocols.
\end{abstract}

\date{\today}
\maketitle

\section{Introduction}
Quantum information science has seen rapid progress in recent years.
Quantum entanglement is one of the driving forces behind quantum information processing, and its generation and distribution is therefore of utmost importance.
Among many other applications, quantum key distribution (QKD) \cite{eker1991, gisin2002, xu2020}, quantum computation \cite{Steane1998, barz2012} and superdense coding \cite{wang2005} heavily rely on the presence of entanglement.
A key feature of photons is that they can be simultaneously entangled in multiple of their degrees of freedom (DOF), thus creating hyper-entangled states \cite{kwiat1997, barreiro2005, suo2015, deng2017, chapman2019}.
By exploiting hyper-entanglement one can greatly increase the dimensionality of the resulting Hilbert space, effectively increasing the quantum information per transmitted photon \cite{kwiat2011}.

Another way of increasing the transmission rate has been a standard in the classical telecommunication industry for quite some time, namely multiplexing. 
Here, many signals are combined (multiplexed) into one transmission channel and separated (de-multiplexed) at the receiver site.
In the quantum regime, multiplexing can be used to overcome the timing limitations imposed by the timing precision and dead time in the detection module, and hence significantly increase the secure key rate \cite{pseiner2021}.

In recent years, the interest in multiplexing of entangled photons in different DOFs has grown \cite{Vergyris_2019, ortega2021, pseiner2021, wengerowsky2018, kim2022}. Achieving coherence and entanglement in the multiplexed DOFs offers another exciting avenue for high-rate quantum communication: directly encoding in energy-time or spatial DOFs also allows for high-dimensional protocols, which offer increased capacity in high-dimensional QKD protocols \cite{tittel2000, ali-khan2007, martin2017} as well as improved noise resilience \cite{ecker2019, doda2020,hu2020b}, essentially only limited by the spatial or temporal resolution of the detection scheme.

In this article, we present an experiment separated into two parts, in which we prove successful transmission of spatial, energy-time as well as polarization entanglement through more than 400 m of multicore fiber (MCF).
At first we certified hyper-entanglement in polarization and energy-time and combined it with the technique of space-division multiplexing.
The entangled photons were spatially multiplexed using a multicore fiber, which consists of 19 single-mode fibers inside a single cladding, connected to a fan-in/fan-out device.
We verified entanglement for four randomly chosen opposite core pairs of the MCF and achieved visibilities up to 94$\%$ in both energy-time and polarization.

In a second step, we verified the transmission of path entanglement through the MCF by non-locally interfering two neighbouring core pairs on two separate beamsplitters.
By varying the phase between the photon pairs, visibility fringes with a visibility up to 96$\%$ were measured. 
As the core pairs are identical, we can conclude that the MCF is indeed capable of transmitting hyper-entanglement in all DOFs faithfully.

The presented approach opens up the possibility of using a single source for generating hyper-entangled photons and efficiently distribute them through a MCF.
Furthermore, because the wavelength of the entangled photons is centered around 1560 nm, the presented approach can be readily integrated into existing telecommunication infrastructure, thus paving the way for higher transmission rates in existing and future QKD protocols as well as other quantum-information applications.
To the best of our knowledge, this is the first demonstration of transmitting a hyper-entangled state in 3 DOFs through a multicore fiber.

\section{Experiment}

\subsection{Energy-time and polarization entanglement analysis}
\label{sec:energy-time}

\begin{figure*}[ht]
\includegraphics[width=1\textwidth]{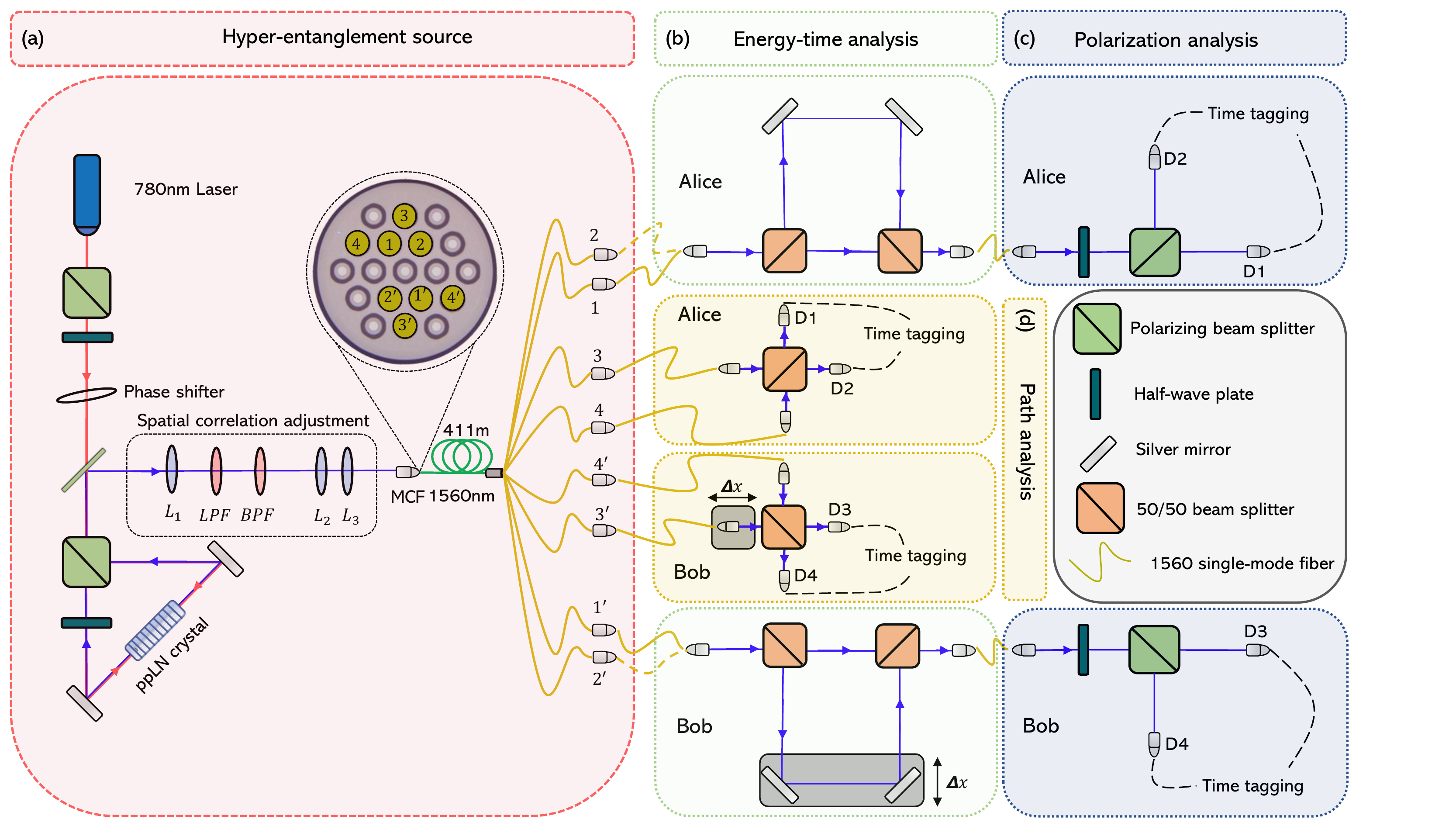}
\caption{Experimental setup. 
\textbf{(a)} Hyper-entanglement source: A MgO:ppLN non-linear crystal is bi-directionally pumped in a Sagnac configuration by a continuous wave laser at 780.24 nm. Because of this special configuration a hyper-entangled state in polarization, energy-time and position-momentum with a center wavelength of 1560.48 nm is produced.
A dichroic mirror (DM) splits the pump light and the SPDC signal. Through the use of a lens configuration the momentum anti-correlations of the entangled photons are imaged onto the front facet of a 19 core MCF. A longpass filter further filters out the pump light while a bandpass filter with a FWHM of $1560 \pm 2.4$ nm selects only the photons centered around the center wavelength. This filtering procedure greatly increases the momentum anti-correlations, since a filtering in wavelength is directly proportional to a filtering on the \textbf{k}-vectors of the photons. Since the opening angle of the emission cone is highly dependent on the crystals' temperature \cite{ortega2022} a temperature of 80.5$\degree$C was chosen for the inner ring (cores 1, 1', 2, 2') while for the outer ring (cores 3, 3', 4, 4') the temperature was 77.5$\degree$C. 
\textbf{(b)} Energy-time analysis: Two opposite core pairs (either 1 and 1' or 2 and 2') are coupled to two separate Franson interferometers. The mirrors of the long arm of Bob's interferometer are mounted on a translation stage. By moving the stage one can vary the phase $\phi$ in Eq. \ref{eq:wavefunction} between the long and short arms.
\textbf{(c)} Polarization analysis: Consisting of a HWP and a PBS. By rotating the HWPs one can specify the measurement basis, either HV or DA.
\textbf{(d)} Path analysis: Two neighbouring core pairs non-locally interfere on two separate 50/50 BS. Through the thusly created path indistinguishability a superposition basis is generated. A key aspect in achieving this is ensuring equal path length differences of all cores inside the coherence length of the entangled photons. By varying the distance of one core to one BS with a translation stage the phase $\theta$ in Eq. \ref{eq:path_wavefunction} is tuned. In order to compensate for polarization drifts in the fiber a PBS was inserted before every input port of the BS, which for simplicity is not shown in the Figure.
}
\label{fig:setup}
\end{figure*}

In the first part of the experiment, we simultaneously studied the entanglement in the energy-time and polarization domain.
The experimental setup consists of a high-dimensional photon pair source and different analysis setups both at Alice's and Bob's site (see Fig.~\ref{fig:setup}). 
The entangled photon pairs were produced via spontaneous parametric down-conversion (SPDC) \cite{walborn2010}.
By pumping a type-0 non-linear SPDC crystal in a Sagnac configuration \cite{kim2006} with a CW-laser at a wavelength of 780.24 nm, polarization-entangled photons with a center wavelength of 1560.48 nm are produced.
The created photons are inherently entangled in the energy-time and position-momentum domain.
Because of the special Sagnac configuration the produced photons also exhibit polarization entanglement.
The photon-pair spectrum is filtered by a $\pm$2.4 nm (FWHM) bandpass filter to select degenerate photons.
By decreasing the spectral width of the photons, the momentum correlations are greatly enhanced in return, which limits the  crosstalk between different cores inside the MCF \cite{ortega2021}. 

Because of momentum conversation in the SPDC process, the signal and idler photons are emitted with opposite transverse momenta, resulting in $\textbf{q}_s = - \textbf{q}_i$, where $\textbf{q}_s$ ($\textbf{q}_i$) refers to the transverse momentum of the signal (idler) photon.
Entanglement in position-momentum for this particular source has already been shown in a previous work \cite{achatz_2022}.
By utilizing a lens system in the far-field plane of the crystal, we imaged the momentum-correlations onto the front facet of a 411-m-long MCF.
Our MCF consists of 19 single-mode fibers (SMF) in a hexagonal pattern (see Figure \ref{fig:setup}).
Because of the anti-correlation in momentum, the partner photon of a collected photon in any core can be found in the diametrically opposite core.
Collecting the photons in opposing cores enables the separation of the two entangled photons despite its wavelength-degenerate spectrum.
For a more detailed description of the lens setup and MCF the reader is referred to \cite{ortega2021}.

Entanglement in the energy-time DOF arises from energy conservation in the SPDC process. 
After the fan-out of the MCF, the entangled photon-pairs are transmitted to two imbalanced Mach-Zehnder interferometers, one for each user, which constitutes a Franson interferometer \cite{franson1989}.
At the first beamsplitter of the interferometer, the photons are either transmitted into the short arm or reflected into the long arm. The propagation difference between the long and the short arm amounts to a delay of 1.2 ns.
They are then recombined at a second beamsplitter.
The cases where both photons take the same paths in both Mach-Zehnder interferometers (either short-short or long-long) are temporally indistinguishable.
Therefore, post-selection on the coincidences reveals non-local interference  \cite{kwiat1993}.
It is important to note that the path length difference $\Delta L$ between the short and long arms of both interferometers must be smaller than the coherence time $\tau_{cc}$ of the entangled photons.
The created hyper-entangled wave function in each core is close to
\begin{equation}
    \ket{\Psi_{E,P}} = \frac{1}{2}\left[\left(\ket{S}\ket{S} + e^{i\phi} \ket{L}\ket{L}\right) \otimes \left(\ket{H}\ket{H} + \ket{V}\ket{V}\right)\right]
\label{eq:wavefunction}
\end{equation}
where $\ket{S}$ and $\ket{L}$ refer to short and long paths and $\ket{H}$ and $\ket{V}$ refer to horizontal and vertical polarization.
The phase $\phi$ can be adjusted via a piezo actuator in the long arm of Bob's interferometer.

After the energy-time analysis, the photons are transmitted to a polarization analysis module, consisting of a half-wave plate and a polarizing beamsplitter. 
They are subsequently detected by superconducting nanowire single photon detectors (SNSPD) with an efficiency of $\sim80\%$ and dark-count rates of $\sim10^2$ Hz. 
On average, the single-photon count rate was about 125 kHz and the coincidence count rate was about 300 Hz.
Accidental coincidences in the order of 5 Hz were subtracted from all coincidence counts.
The low heralding efficiency, i.e. coincidence-to-single-ratio, is a result of the low-efficiency coupling of the SPDC light into the MCF. 
A micro-lens array \cite{Dietrich2017} could greatly increase the heralding efficiency of our source.

\subsection{Path entanglement analysis}
\label{sec:path}

As a second step the path entanglement transmitted through the MCF was analyzed.
The experimental setup for the source is identical to the one explained in Sec. \ref{sec:energy-time}.
In order to  analyze path entanglement two core pairs in the outer ring of the MCF were non-locally interfered on two BS, as is shown in Fig. \ref{fig:setup}.
By guaranteeing an equal distance  from the front facet of the MCF to the interfering point on the BS a superposition basis is created \cite{rossi2008}. 
The wave function projected on these two cores would ideally read as:

\begin{equation}
    \ket{\psi_S} = \frac{1}{\sqrt{2}}\left(\ket{3}\ket{3'} - e^{i\theta}\ket{4}\ket{4'}\right)
\label{eq:path_wavefunction}
\end{equation}

where $\ket{3}$, $\ket{3'}$, $\ket{4}$ and $\ket{4'}$ refer to the different cores of the MCF.
Importantly, the path length difference between all cores has to be smaller than the coherence length of the entangled photons.
By adjusting the path length difference of one core with a piezo electrical actuator, effectively varying the phase $\theta$, visibility fringes between different output ports of the BS are observed. 
Since only photons with the same polarization state interfere, a PBS was inserted in front of the BS.
By this measure polarization drifts in one core will not affect the path visibility but rather just reduce the heralding. 
Accidental coincidences in the order of 10 Hz were subtracted from all coincidence counts.

It is important to note that the measurements from \ref{sec:energy-time} and \ref{sec:path} were done sequentially since only 4 detector channels were available. 
With enough detectors one can, in principle, measure energy-time, polarization and path entanglement in parallel, opening up the possibility of using just one source to efficiently generate and distribute entanglement in multiple DOFs between different users.

Before we move on to results let us briefly define the overall target state, which we would expect to feed into the MCF and hope to certify at the output:
\begin{align}
    \ket{\Psi_{T}}:=\frac{1}{4}(\ket{11'}+\ket{22'}+\ket{33'}+\ket{44'})\otimes \nonumber\\(\ket{SS} + \ket{LL}) \otimes (\ket{HH} + \ket{VV})
\end{align}
\section{Results}

\subsection{Energy-time and polarization visibility measurements}

\begin{figure}
\includegraphics[width=0.45\textwidth]{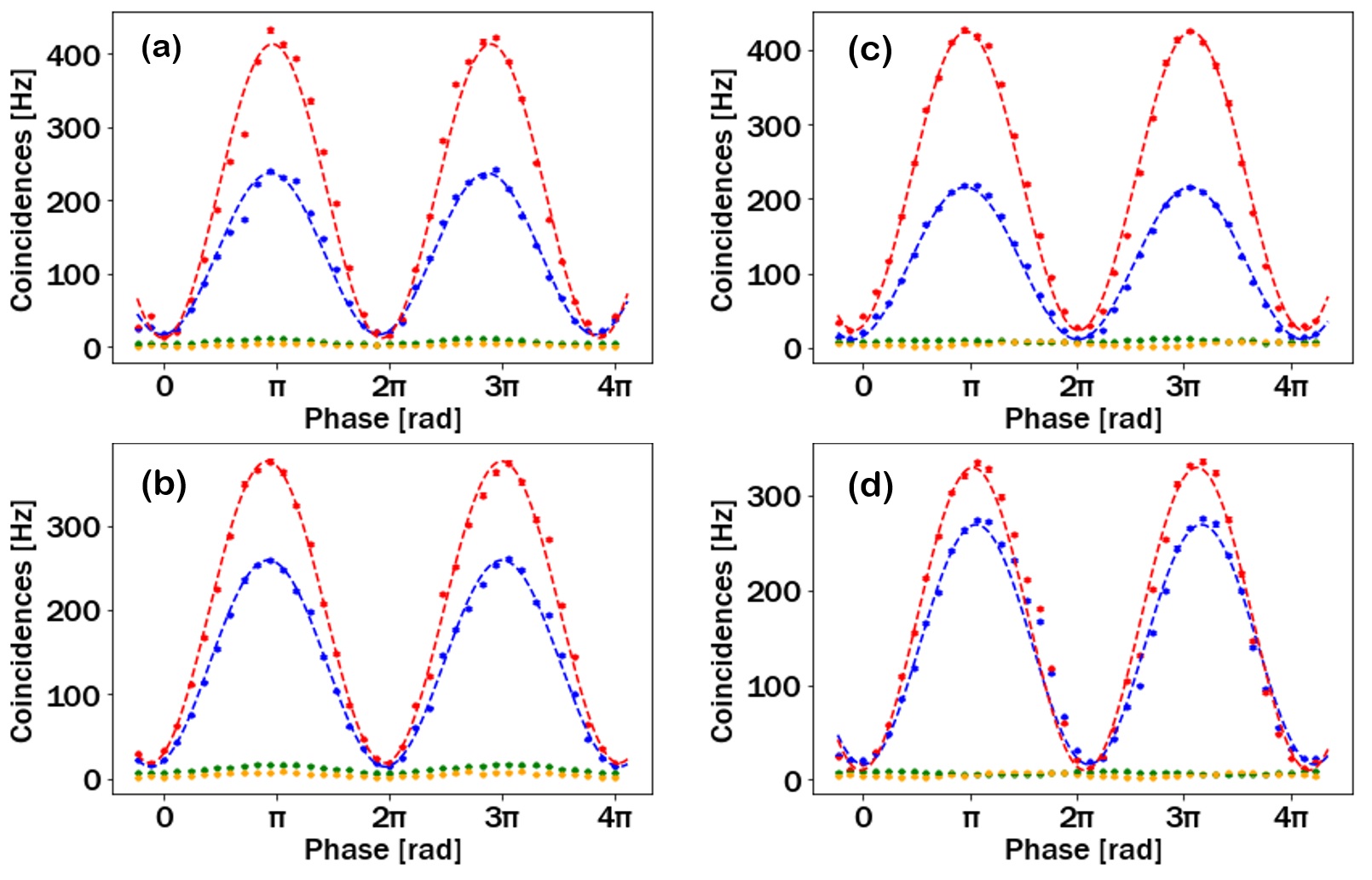} 
    \caption{Measured time visibility fringes by varying Bob's phase. Blue represents the coincidences for HH (DD), red for VV (AA), green for HV (DA) and yellow for VH (AD). \textbf{(a)} Core pair 1-1' HV Basis, \textbf{(b)} Core pair 1-1' DA Basis, \textbf{(c)} Core pair 2-2' HV Basis, \textbf{(d)} Core pair 2-2' DA Basis. For each data point the integration time was 30s. We assumed Poissonian statistics, note that the errors are smaller than the data points.
    }
    \label{fig:fringes} 
\end{figure}

\begin{figure}
\includegraphics[width=0.45\textwidth]{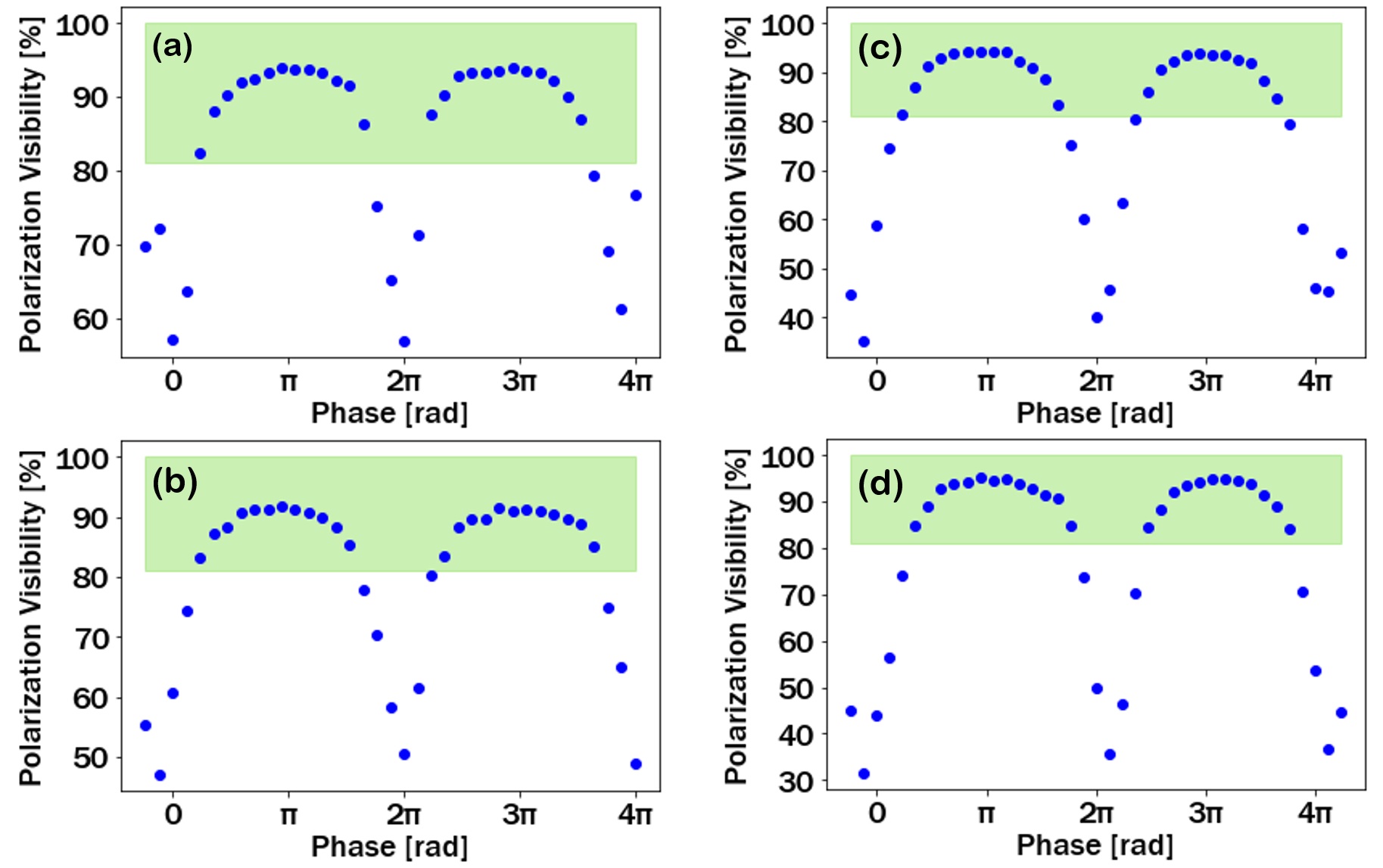} 
    \caption{Measured polarization visibility fringes by varying Bob's phase. The green area represents a visibility of greater than 81$\%$, representing the limit for QKD protocols. \textbf{(a)} Core pair 1-1' HV Basis, \textbf{(b)} Core pair 1-1' DA Basis, \textbf{(c)} Core pair 2-2' HV Basis, \textbf{(d)} Core pair 2-2' DA Basis. At multiples of $2\pi$ the polarization visibility drops. This stems from the fact that destructive interference for the coincidences occurs, thus fewer coincidences are measured. For each data point the integration time was 30s. We assumed Poissonian statistics, note that the errors are smaller than the data points.
    }
    \label{fig:pol_vis} 
\end{figure}

In order to verify the transmission of hyper-entanglement through the MCF, at first, both HWPs at the polarization modules were set to H, inducing no phase shift on the polarization state of the photons, thus corresponding to a measurement in the HV-basis.
To analyze the hyper-entangled state in its entirety,  the long arm of Bob's interferometer was scanned, effectively varying the phase $\phi$ in the first part of Eq. \ref{eq:wavefunction}.
The obtained visibility fringes are shown in Fig. \ref{fig:fringes}.
At every step $\Delta x$ of Bob's interferometer the integration time was $30s$ and the coincidence counts for the detector combinations HH (D1/D3), VV (D2/D4), HV (D1/D4) and VH (D2/D3) were recorded.
Thereby a polarization visibility at every $\Delta x$ is obtained.
The polarization visibilities are shown in Fig. \ref{fig:pol_vis}, where the green area represents a visibility of higher than 81$\%$.
Since a $\ket{\Phi^+}$ Bell state was produced, correlations only arise between HH (DD) and VV (AA), while there are no correlations between HV (DA) and VH (AD).
The polarization visibility is calculated with
\begin{equation}
    V_{pol} = \frac{CC_{HH} + CC_{VV} - CC_{HV} - CC_{VH}}{CC_{HH} + CC_{VV} + CC_{HV} + CC_{VH}},
\end{equation}
where $CC_{signal,idler}$ corresponds to the coincidence counts for each output port of the polarizing beam splitter.
In order to calculate the visibilities in the energy-time DOF the obtained fringes were fitted with a sine function and calculated with 
\begin{equation}
    V_{time} = \frac{CC_{max} - CC_{min}}{CC_{max} + CC_{min}}.
\end{equation}
The same procedure was repeated for a measurement in the DA basis, meaning that both HWPs in the polarization modules were rotated by 22.5\degree.
The results of the visibility fringe measurements for different polarization settings and cores are shown in Tables \ref{tab:visibility1} and \ref{tab:visibility2}.

\begin{table}[h]
    \begin{center}
    \begin{tabular}{c||c} 
     \textbf{Detector combination} & \textbf{Time visibility [$\%$]} \\ [0.5ex] 
     \hline\hline \\
     HH & 86.6 $\pm$ 3.1  \\   [1ex]
    
     VV & 94.2 $\pm$ 1.6  \\  [1ex]
    
     DD & 89.5 $\pm$ 2.7  \\  [1ex]
    
     AA & 90.6 $\pm$ 2.1  \\ \\
    \end{tabular}
    \caption{Obtained visibilities from scanning the long arm ob Bob's interferometer for the core combination 1-1'. HH and VV correspond to measurements in the HV basis while DD and AA correspond to measurements in the DA basis.}
    \label{tab:visibility1}
    \end{center}
\end{table}

\begin{table}[h]
    \begin{center}
    \begin{tabular}{c||c}
     \textbf{Detector combination} & \textbf{Time visibility [$\%$]} \\ [0.5ex] 
     \hline\hline\\
     HH & 90.1 $\pm$ 2.9  \\  [1ex]
    
     VV & 89.5 $\pm$ 2.1  \\  [1ex]
    
     DD & 88.7 $\pm$ 2.7  \\  [1ex]
    
     AA & 94.1 $\pm$ 1.8 \\ \\
    \end{tabular}
    \caption{Obtained visibilities from scanning the long arm ob Bob's interferometer for the core combination 2-2'. HH and VV correspond to measurements in the HV basis while DD and AA correspond to measurements in the DA basis.}
    \label{tab:visibility2}
    \end{center}
\end{table}

As can be seen in Table \ref{tab:visibility1} and \ref{tab:visibility2} all visibilities are greater than 81$\%$, which is the limit for QKD protocols for which a key can be extracted.


\subsection{Path visibility measurements}

\begin{figure}
\includegraphics[width=0.45\textwidth]{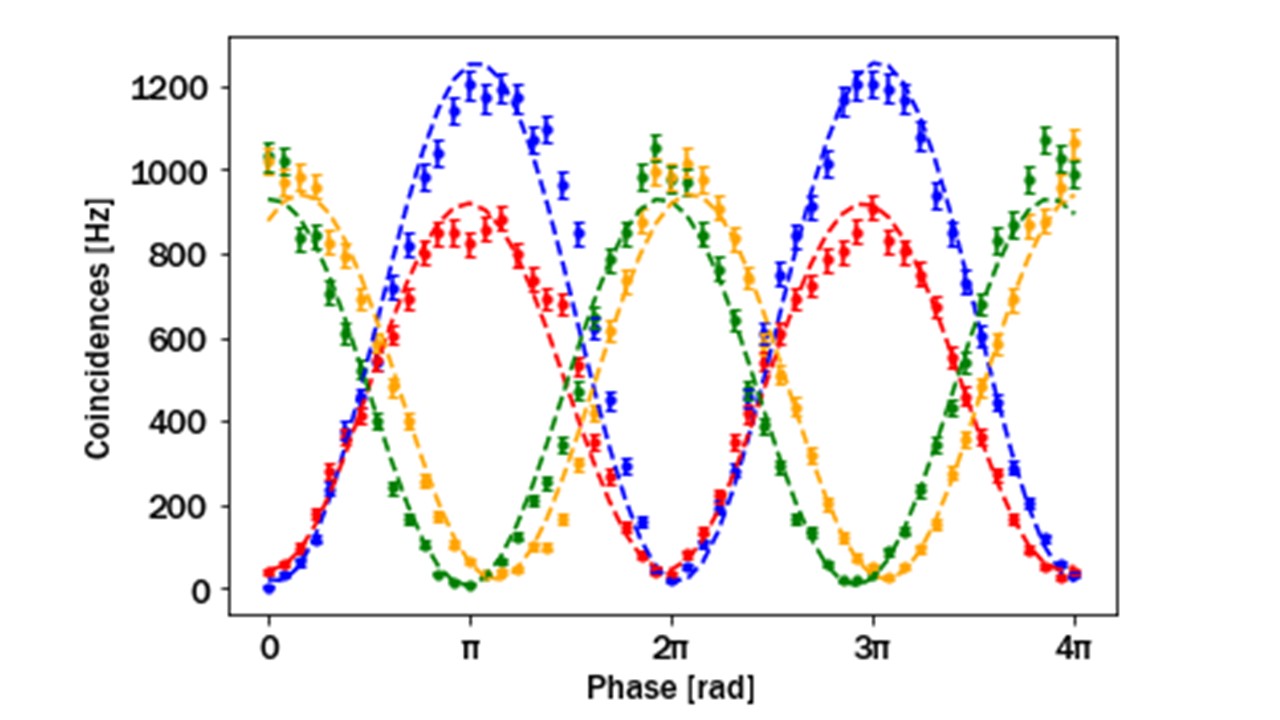} 
    \caption{Measured path visibility fringes between different detector combinations with an integration time of 1s. The detector combinations are the following: blue D1/D3, red D2/D4, green D1/D4 and yellow D2/D3. For the errors we assumed Poissonian statistics.
    }
    \label{fig:path_vis} 
\end{figure}

In order to verify the transmission of path entanglement through the MCF the path lengths of all cores have to be identical in the order of the coherence length of the photons.
By varying the distance to the BS of one core via a piezo electrical actuator the phase $\theta$ in Eq. \ref{eq:path_wavefunction} is tuned.
Coincidence counts between all 4 detector combinations were recorded with an integration time of 1 s.
The obtained results are shown in Fig. \ref{fig:path_vis}.
In order to obtain the visibilities the obtained fringes were fitted with a sine function and subsequently calculated with 
\begin{equation}
    V_{path} = \frac{CC_{max} - CC_{min}}{CC_{max} + CC_{min}}
\end{equation}
The obtained visibilities are listed in Tab. \ref{tab:path_vis}. 
As can be seen visibilities as high as 97.6$\%$ were achieved, demonstrating a high degree of entanglement.

\begin{table}[h]
    \begin{center}
    \begin{tabular}{c||c}
     \textbf{Phase} & \textbf{Path visibility [$\%$]} \\ [0.5ex] 
     \hline\hline\\
     0 & 97.6 $\pm$ 0.6  \\  [1ex]

     $\pi$ & 95.8 $\pm$ 0.9 \\ \\
    \end{tabular}
    \caption{Obtained visibilities from varying the length of one core. The two points correspond to visibilities for different measurement bases, which we took to define $\sigma_x\otimes\sigma_x$ and $\sigma_y\otimes\sigma_y$.}
    \label{tab:path_vis}
    \end{center}
\end{table}

In Fig. \ref{fig:path_vis} a slight offset between different detector combinations is noticeable. 
This most likely stems from a slight length-offset between the non-interfering core pairs.
This fact, however, does not compromise the visibility results.

The visibility implies a verified off-diagonal element in the path basis of $|\bra{33'}\rho\ket{44'}|=0.24$ \cite{hu2020}. In this non-adversarial scenario one can safely assume that this value is representative for all the elements $|\bra{ii'}\rho\ket{jj'}|$. This, together with the coincidences across the cores, i.e. the measured diagonal basis elements in path basis, gives a final fidelity with the maximally path entangled state of $\mathrm F_{path}=0.953$, which is well above the bound for qutrit entanglement of $0.75$, thus proving a Schmidt number of four or in other words genuine four-dimensional entanglement.

\section{Discussion}

The presented approach offers multiple advantages for applications which need to be highlighted.
While we connected separate measurement setups to verify entanglement in different DOFs, the source (except for crystal temperature) and the fiber remained unperturbed, proving simultaneous presence of all hyper-entangled DOFs after MCF transmission. 
First, we demonstrated the successful transmission of a spatially multiplexed polarization and energy-time hyper-entangled state through the inner cores of the 411-m-long MCF. 
By using the technique of multiplexing, timing limitations in present detector systems are circumvented and the secure key rate is increased \cite{pseiner2021}.
Since the photons' momenta are anti-correlated, one can distribute multiple pairs to different users by using opposing cores of a multicore fiber.
Furthermore by using two MCFs, one for each user, one can use the intrinsic phase-stability of MCFs \cite{bacco2020} for efficiently distributing high-dimensional entanglement.
One can also use the spatial entanglement across cores to maximise rates for a single pair of users. 
In this context, and as a second step, we verified entanglement in the path DOF by non-locally interfering two opposite core pairs of the outer ring of the MCF on two beamsplitters.
We achieved visibilities as high as 97$\%$, underlining the high-dimensional nature of the path entanglement. 
Another possible application is superdense coding \cite{barreiro2008} which has recently been realized with qu-quarts in Ref.~\cite{min2018} as well as superdense teleportation \cite{graham2015} and single-copy entanglement distillation \cite{Ecker2021}.
Since our entanglement source naturally produces wavelength-correlated photon pairs, the presented approach offers high scalability by additionally integrating dense wavelength-division multiplexing, as is done for example in Ref.~\cite{Vergyris_2019}.
The used wavelength of 1560 nm makes our approach readily integrable into current telecommunication infrastructure, thus paving the way for highly efficient quantum information protocols over deployed fiber networks.


\section*{Acknowledgments}
We gratefully acknowledge financial support from the Austrian Academy of Sciences and the EU project OpenQKD (Grant agreement ID: 85715).
E.A.O. acknowledges ANID for the financial support (Becas de doctorado en el extranjero “Becas Chile”/2016 – No. 72170402).
M.H. would like to acknowledge funding from the European Research Council (Consolidator grant 'Cocoquest' 101043705)

\bibliography{biblio}

\end{document}